\documentclass[twocolumn,aps,superscriptaddress,showpacs,nofootinbib,floatfix]{revtex4-1}
\usepackage{epsfig,bm}
\usepackage{graphicx,amssymb}
\usepackage{amsmath}
\usepackage{physics}
\usepackage{tensor}
\usepackage{diagbox}
\usepackage{epstopdf}
\usepackage{float}
\usepackage{tabularx}
\usepackage[eps]{pstricks}
\usepackage[normalem]{ulem}  

\newcommand{\pt}{$\mathrm{p_{T}}$}

\newcommand{\pc}{$P_{c}$}
\newcommand{\jpsi}{$J /\psi$}

\newcommand{\sqrts}{$\sqrt{s}$}

\renewcommand\sout{\bgroup \color{red} \ULdepth=-.5ex \ULset}
\usepackage{slashed}
\newcommand{\comment}[1]{}

\begin{document}

\title{Production of $P_c(4312)$ state in electron-proton collisions}


\author{In Woo Park}%
\affiliation{Department of Physics and Institute of Physics and Applied Physics, Yonsei University, Seoul 03722, Korea}

\author{Sungtae Cho}
\affiliation{Division of Science Education, Kangwon National
University, Chuncheon 24341, Korea}
\affiliation{Center for Extreme Nuclear Matters (CENuM), Korea University, Seoul, Korea}
\author{Yongsun Kim}%
\email{yongsun.kim@cern.ch}
\affiliation{Department of Physics, Sejong University, Seoul, Korea}
\affiliation{Center for Extreme Nuclear Matters (CENuM), Korea University, Seoul, Korea}

\author{Su Houng Lee}%
\email{suhoung@yonsei.ac.kr}
\affiliation{Department of Physics and Institute of Physics and Applied Physics, Yonsei University, Seoul 03722, Korea}


\begin{abstract}
We study the cross sections for the electro-production of $P_c(4312)$ particle, a recently discovered pentaquark state, in electron-proton collisions assuming possible quantum numbers to be  $J^{P}=\frac{1}{2}^\pm, \frac{3}{2}^\pm$.  $\sqrt{s}$ is set to the energy of the future Electron Ion Collider at Brookhaven National Laboratory, in order to asses the possibility of the measurement in this facility.
One can discriminate the spin of $P_c(4312)$ by comparing the pseudorapidity distribution in two different polarization configurations for proton and electron beams. Furthermore, the parity of $P_c(4312)$ can be discerned by analyzing the decay angle in the $P_c \rightarrow p +J/\psi$ channel. As the multiplicity of $P_c$ production in our calculation is large, the EIC can be considered as a future facility for precision measurement of heavy pentaquarks.
\end{abstract}


\maketitle

\section{Introduction}

Recent years have witnessed the observation of a series of
pentaquark state candidates from the measurements at the Large
Hadron Collider (LHC); the first observation of probable pentaquark states $P_c(4380)$ and $P_c(4450)$ was reported by LHCb
collaboration in 2015 \cite{LHCb:2015yax}, and later the
observation of  \pc(4312), \pc(4440), and \pc(4457) was made in 2019 \cite{LHCb:2019kea}.
The $P_c(4450)$ measured earlier in 2015 was confirmed, but revealed to consist of 
two narrow overlapping peaks \pc(4440), and \pc(4457) by the investigation of $J /\psi+p$ decays in $pp$ collisions at \sqrts~= 7, 8 and 13 TeV.
More recently, the LHCb collaboration discovered a strange pentaquark
state $P_{cs}(4458)$ in the $J/\psi \Lambda$ invariant mass
distribution from an analysis of the $\Xi_b^- \rightarrow
    J/\psi + \Lambda + K^-$ decay channel \cite{LHCb:2020jpq}.

These heavy pentaquark states confirmed the existence
of exotic hadrons and inspired a diverse discussion about their
internal structures and the quantum numbers; Are they in molecular configurations 
or compact multiquark states?   Do just kinematical effects generate these resonances?~\cite{Chen:2016qju, Liu:2019zoy}  
What are the spins and parities of them? There have been several theoretical approaches 
to answer these questions, including quark models, meson-based models, diquark-based models, and QCD sum rules, yet without making no consensus.

Given the observation of the $P_c$ decay into $J/\psi$ and proton, we can expect to create the $P_c$ by colliding a proton with a photon which  couples to $J/\psi$.  Thereby, we propose electron-proton $(e+p)$ collision experiment to create statistically meaningful \pc~states, thus providing critical evidence for their quantum numbers. One of the standard methods to determine the spins and parity of an unknown particle is to examine their angular distribution. Moreover, it would be beneficial if that experiment could adjust the spin polarity of colliding particles. In that sense, the e+p collision with polarized beams will provide desirable circumstances.
 
The Electron Ion Collider (EIC) is a future collider to be built at
Brookhaven National Laboratory (BNL) \cite{eicYellow} which is
designed to collide an electron beam with proton, deuteron and
various heavy ion beams at high luminosity. The EIC can be a great
factory for the $P_c$ production.
A large coverage of detector
system  will be useful to measure \pc~ $\rightarrow p + J/\psi \rightarrow p + e^{+} + e^{-}$. Two
prospective experiments proposed at the EIC, ECCE \cite{ecce} and ATHENA \cite{athena}, meet this requirement well.

In this paper, we study the angular distribution of $P_c$(4312) production at the EIC's design energy $\sqrt{s} =$
126 GeV ($\mathrm{E}_{e}$ = 16 GeV and $\mathrm{E}_{p}$ = 250 GeV). The differential cross sections are formulated for possible combinations of spin and parity. For the technical evaluation, we use the vector meson dominance (VMD) approach.  The interaction strength is derived from the decay width of $P_c$(4312) measured by the LHCb collaboraiton.

This paper is organized as follows. In Section II, we introduce
the VMD model to determine the coupling strength of a proton, a $\gamma$, and a $P_c$(4312).
In Section III, we calculate the cross section of \pc~ production under four situation of spin($\frac{1}{2}$ or $\frac{3}{2}$) and parity($\pm$).
In Section IV, the analysis of differential cross section is presented. The last section is given for the summary.

\section{Coupling strength~: ~$g_{\gamma p P_c}$}

\begin{figure}[h]
\includegraphics[width=2.5in,height=1.25in]{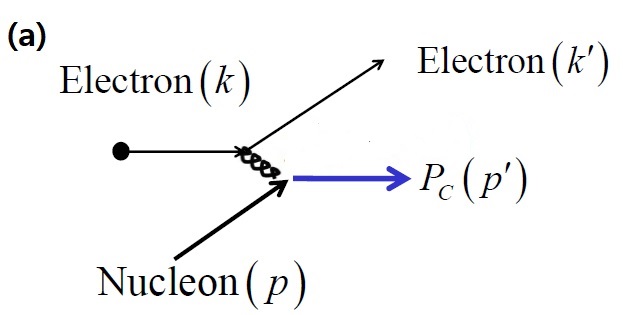}
\includegraphics[width=2.5in,height=1.35in]{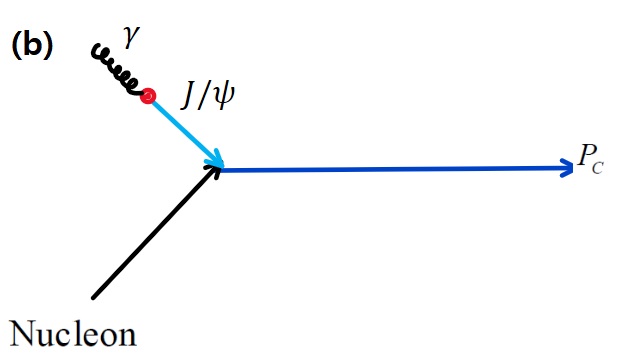}
\caption{  (a) The electro-production of a pentaquark on the proton
target. The effective proton-$\gamma$-pentaquark coupling is
described in the VMD framework.  (b) the coupling between a proton, a $\gamma$, and a pentaquark is mediated by the $J/\psi$ meson in the VMD model.}
\label{fig:electro-production}
\end{figure}

We consider the pentaquark which is electro-produced from a
proton target; $P_c$ is produced by the interaction between
the proton and a photon ($\gamma$) emitted from the electron. Fig.
\ref{fig:electro-production}(a) describes the process to the leading order with an effective coupling strength $g_{\gamma p P_c}$ between a
proton, a $\gamma$, and a pentaquark. Although our calculation is
carried out only for the $P_c(4312)$ in this paper, it can be generalized to other pentaquark states.

To compute the coupling strength $g_{\gamma p P_{c}}$, we use the VMD hypothesis and assume that the experimental estimate of the $P_c(4312)$ width (9.8 MeV)~\cite{LHCb:2019kea} is dominated by its
$P_c \rightarrow p+J/\psi$ decay. This approximation provides an upper bound for $g_{\gamma p P_c}$ because all the measured pentaquark states could in principle also
decay into a charmed baryon and meson such as $P_c \rightarrow \Lambda_c
+\bar{D}$.  

\subsection{Coupling between $J/\psi$, $p$, and $P_c$: $g_{J p P_c}$}
The VMD model states that photon interacts with hadrons through vector mesons as shown in \ref{fig:electro-production}(b). In the $P_{c}$-creating channels, $J/\psi$ acts as the main player because it contains a $c\bar{c}$ pair \cite{Klingl:1996by}. Therefore, the first step is to determine the coupling between
$P_c$, $J/\psi$, and $p$, called $g_{J p P_c}$.
The form of interaction depends on
the quantum numbers of $P_c$, and we
choose the following derivative effective Lagrangians
depending on the spin-parity ($J^P$) state.

\begin{align}
{\cal L}_{\rm int} = \begin{cases}
\frac{g_{JpP_{c}}}{m_{J/\psi}}\bar{\psi}_p\sigma^{\mu\nu}F_{\mu\nu}^J\psi_{P_c}
&~~J^P=\frac{1}{2}^{+}, \cr
\frac{g_{JpP_{c}}}{m_{J/\psi}}\bar{\psi}_p\gamma_{5}\sigma^{\mu\nu}F_{\mu\nu}^J\psi_{P_c}
&~~J^P=\frac{1}{2}^{-}, \cr
\frac{g_{JpP_{c}}}{m_{J/\psi}}\bar{\psi}_p\gamma_{5}\gamma^{\mu}F_{\mu\nu}^J\psi^{\nu}_{P_c}
&~~J^P=\frac{3}{2}^{+}, \cr
\frac{g_{JpP_{c}}}{m_{J/\psi}}\bar{\psi}_p\gamma^{\mu}F_{\mu\nu}^J\psi^{\nu}_{P_c}
&~~J^P=\frac{3}{2}^{-}. \cr
\end{cases}
\label{eq:Lagrangian}
\end{align}
, where $\psi_p$, $A_\mu^J$, and $\psi_{P_c}$ are the fields of proton,
$J/\psi$, and $P_c$, respectively. We also use the convention,
$F_{\mu\nu}^J=\partial_\mu A_\nu^J-\partial_\nu A_\mu^J$,
$\sigma^{\mu\nu}=(\gamma^\mu\gamma^\nu-\gamma^\nu \gamma^\mu)/2$,
with the gamma matrices, $\gamma^\mu$.



Based on Eq. (\ref{eq:Lagrangian}), the decay width can be calculated as

\begin{eqnarray}
&\Gamma_{P_c\to p+J/\psi}=\frac{1}{8\pi}\frac{|\vec{p_{f}}|}{m_{P_{c}}^{2}}|\mathcal{M}|^{2}
\label{eq:couplingStrength}
\end{eqnarray}

with $\mathcal{M}$ being the invariant matrix amplitude, and $\vec{p_{f}}$ being the momentum of the decayed
particle in the center of mass (CM) frame: we summarize relevant formulas in Appendix
\ref{appendix:jppc}. The masses of $P_c$(4312) and $J/\psi$ are taken from the Particle Data Group~\cite{Zyla:2020zbs}: $m_{P_{c}}=4311.9~\text{MeV}$, $m_{J/\psi}=3096.9~\text{MeV}$.
By equating Eq.~\eqref{eq:couplingStrength} with the LHCb result, we can derive $g_{JpP_{c}}$ as summarized in Table.~\ref{table:coupling}.
\begin{table}[H]
\caption{The interaction strength $g_{JpP_c}$ between a $P_c$,
a $p$ and a $J/\psi$ in the VMD model}
\centering
 \begin{tabular}{||c c c c c c||}
 \hline
  $J^P$  &  $\frac{1}{2}^+$ & $\frac{1}{2}^-$ & $\frac{3}{2}^+$  & $\frac{3}{2}^-$ &
  \\ [0.5ex]
   \hline\hline
$~g_{JpP_c} $ ~&~ 0.379 ~&~ 0.169 ~&~ 1.47 ~&~ 0.599 ~&
 \\ \hline
\end{tabular}
\label{table:coupling}
\end{table}

\subsection{Coupling between $J/\psi$ and $\gamma$: $g_{J}$}
Regarding $J/\psi \rightarrow e^{-}+e^{+}$, we adopt the following interaction Lagrangians for $J/\psi$-$\gamma$
and $\gamma$-dilepton interactions, respectively,

\begin{eqnarray}\label{eqn:photo-J}
 \mathcal{L}_{J/\psi\gamma} & = & -\frac{e}{2g_{J}}F^{\mu\nu}F^{J}_{\mu\nu},
\nonumber \\
 \mathcal{L}_{\gamma e^{-} e^{+}}& =&
-e\bar{\psi}\gamma^{\mu}A_{\mu}\psi.
\end{eqnarray}
where $g_{J}$ is the coupling constant between the $J/\psi$ and the $\gamma$. 
Using the invariant matrix element given in
Appendix \ref{appendix:jpsi}, we can relate $g_{J}$ to 
the decay width of $J/\psi \rightarrow  e^{-}+e^{+}$:

\begin{align*}
\Gamma &=\frac{4\pi}{3}\frac{\alpha^2}{g_J^2}\sqrt{m_{J/\psi}^{2}
-4m_{l}^{2}}(1+\frac{2m_{l}^{2}}{m_{J/\psi}^{2}}) \\
&=92.9~\text{keV}\times0.05971,
\end{align*}
, from which we obtain $g_{J}$=11.2.

\subsection{Relationship between $g_{JpP_{c}}$, $g_{\gamma P_{c}}$, and  $g_{J}$}
Finally, we can derive $g_{\gamma P_{c}}$ from $g_{JpP_{c}}$ and  $g_{J}$ using the Lagrangians given in Eq (\ref{eqn:photo-J}).
\begin{align}
g_{\gamma pP_{c}}=-\frac{eg_{JpP_{c}}q^{2}}{g_{J}}\frac{1}
{q^{2}-m_{J/\psi}^{2}}. \label{caseI}
\end{align}
where $q$ is the momentum of the $J/\psi$.


\begin{widetext}
\section{Cross section calculation}

In this section, we calculate the invariant amplitudes for the
production of $P_c$ state in four possible spin-parity situation. The
relevant diagram is given in Fig. \ref{fig:electro-production}(a).

\subsection{Cross sections with unpolarized beams}

Invariant matrix amplitudes for each Lagrangian shown in Eq.
\eqref{eq:Lagrangian} are given by,

\begin{align}
\mathcal{M} = \begin{cases} \frac{eg_{\gamma
pP_c}}{m_{J/\psi}}\bar{u}^{l'}(k')
\gamma^{\nu}u^{l}(k)\frac{2q^{\mu}}{q^2}\bar{u}^{P_c}(p')
\sigma_{\mu\nu}u^{N}(p) &~~J^P=\frac{1}{2}^{+}, \cr
\frac{eg_{\gamma pP_c}}{m_{J/\psi}}\bar{u}^{l'}(k')
\gamma^{\nu}u^{l}(k)\frac{2q^{\mu}}{q^2}\bar{u}^{P_c}(p')
\gamma_5\sigma_{\mu\nu}u^{N}(p) &~~J^P=\frac{1}{2}^{-}, \cr
\frac{eg_{\gamma pP_c}}{m_{J/\psi}}\bar{u}^{l'}(k')
\gamma^{\alpha}u^{l}(k)\frac{(q_{\mu}g_{\alpha\nu}-q_{\nu}
g_{\alpha\mu})}{q^2}\bar{u}^{P_{c}\mu}(p')\gamma_{5}
\gamma^{\nu}u^{N}(p) &~~J^P=\frac{3}{2}^{+}, \cr
 \frac{eg_{\gamma pP_c}}{m_{J/\psi}}\bar{u}^{l'}(k')
\gamma^{\alpha}u^{l}(k)\frac{(q_{\mu}g_{\alpha\nu}-q_{\nu}
g_{\alpha\mu})}{q^2}\bar{u}^{P_{c}\mu}(p')\gamma^{\nu}u^{N}(p)
&~~J^P=\frac{3}{2}^{-}. \cr
\end{cases}
\label{eq:amplitudes}
\end{align}

We sum the square of the results for final spins and take the average of the initial spin polarizations of the incoming  electron and
proton. The detailed computation is shown in Appendix
\ref{appendix:unpolarized amplitude}. The results show that the
differences in the spin-averaged square of the invariant
amplitudes between opposite parities, Eqs. (\ref{half_polized})
and (\ref{threehalves_polized}), appear in the differences in the  sign for the $m_{P_{c}}$ term.

\subsection{Cross sections with polarized beams}
Considering the operation of spin-polarized beams of electron and proton, we also study the polarization dependencies of the
electro-production cross section.  In order to describe polarized
electrons and protons, we use the 
projection operator, $P_{\text{R/L}}=\frac{1\pm\gamma_{5}\slashed{s}}{2}$, which satisfies
\begin{equation}
\frac{1+\gamma_{5}\slashed{s}}{2}u(p,s)=u(p,s), \qquad
\frac{1-\gamma_{5}\slashed{s}}{2}u(p,-s)=u(p,-s),
\end{equation}
with the spin 4-vector, $s^{\prime\mu}=(0,\vec{s'})=(0,\vec{p}/p)$. $\vec{s'}$ 
is the spin polarization vector in the rest frame and $\vec{p}$ is the momentum of polarized particle. The spin
4-vector becomes in the Lorentz transformation,
\begin{equation}
s^{\mu}=(\frac{\vec{p}\cdot\vec{s'}}{m},\vec{s'}+\frac{\vec{p}
\cdot\vec{s'}}{m(E+m)}\vec{p})=(\frac{p}{m},\frac{E\vec{p}}{mp}).
\end{equation}
It results in
\begin{equation}
\sum_{i=1}^{2}u^{i}(p)\bar{u}^{i}(p)\frac{1\pm\gamma_{5}
\slashed{s}}{2}=(\slashed{p}+m)\frac{1\pm\gamma_{5}\slashed{s}}{2}
=\frac{\slashed{p}+m\pm\slashed{p}\gamma_{5}\slashed{s}\pm
m\gamma_{5} \slashed{s}}{2}. \label{spin_projection}
\end{equation}
In high energy limit, ($m\to 0$), ~$s^{\mu}\approx p^{\mu}/m$, Eq.
(\ref{spin_projection}) becomes, $\slashed{p}
\frac{1\pm\gamma_{5}}{2}$, and therefore we can approximate the
projection operator for massless particles, or electrons, as
$P_{R/L}=\frac{1\pm\gamma_{5}}{2}$.

With the above spin projection operator, we consider the invariant
amplitudes for the cross sections with polarized electrons and
protons. Here, electrons and protons are chosen to be $RR$, $RL$,
$LR$, $LL$, where $R$ and $L$ represent the right-handed and the left-handed
polarity, respectively.

\begin{align}
\mathcal{M} = \begin{cases} \frac{2eg_{\gamma
PP_c}}{m_{J/\psi}q^2}\bar{u}^{l'}(k')
\gamma^{\beta}\frac{1\pm\gamma_{5}}{2}u^{l}(k)\bar{u}^{P_c}(p')
q^{\nu}\sigma_{\nu\beta}\frac{1\pm\gamma_{5}\slashed{s_p}}{2}
u^{N}(p) &~~J^P=\frac{1}{2}^{+} RR,LL, \cr \frac{2eg_{\gamma
PP_c}}{m_{J/\psi}q^2}\bar{u}^{l'}(k')\gamma^{\beta}
\frac{1\pm\gamma_{5}}{2}u^{l}(k)\bar{u}^{P_c}(p')q^{\nu}
\sigma_{\nu\beta}\frac{1\mp\gamma_{5}\slashed{s_p}}{2}u^{N}(p)
&~~J^P=\frac{1}{2}^{+} RL,LR,\\
\frac{2eg_{\gamma PP_c}}{m_{J/\psi}q^2}\bar{u}^{l'}(k')\gamma^{\beta}
\frac{1\pm\gamma_{5}}{2}u^{l}(k)\bar{u}^{P_c}(p')\gamma_{5}
q^{\nu}\sigma_{\nu\beta}\frac{1\pm\gamma_{5}\slashed{s_p}}{2}
u^{N}(p) &~~J^P=\frac{1}{2}^{-} RR,LL,\\
\frac{2eg_{\gamma PP_c}}{m_{J/\psi}q^2}\bar{u}^{l'}(k')
\gamma^{\beta}\frac{1\pm\gamma_{5}}{2}u^{l}(k)\bar{u}^{P_c}(p')
\gamma_{5}q^{\nu}\sigma_{\nu\beta}\frac{1\mp\gamma_{5}
\slashed{s_p}}{2}u^{N}(p) &~~J^P=\frac{1}{2}^{-} RL,LR,\\
\frac{eg_{\gamma PP_c}}{m_{J/\psi}q^2}\bar{u}^{l'}(k')\gamma^{\beta}\frac{1\pm\gamma_{5}}{2}u^{l}(k)\bar{u}^{\mu}(p')\gamma_{5}\gamma^{\nu}(q_{\mu}g_{\beta\nu}-q_{\nu}g_{\beta\mu})\frac{1\pm\gamma_{5}\slashed{s_{p}}}{2}u^{N}(p)&~~J^P=\frac{3}{2}^{+} RR,LL,\\
\frac{eg_{\gamma PP_c}}{m_{J/\psi}q^2}\bar{u}^{l'}(k')\gamma^{\beta}\frac{1\pm\gamma_{5}}{2}u^{l}(k)\bar{u}^{\mu}(p')\gamma_{5}\gamma^{\nu}(q_{\mu}g_{\beta\nu}-q_{\nu}g_{\beta\mu})\frac{1\mp\gamma_{5}\slashed{s_{p}}}{2}u^{N}(p)&~~J^P=\frac{3}{2}^{+} RL,LR,\\
\frac{eg_{\gamma PP_c}}{m_{J/\psi}q^2}\bar{u}^{l'}(k')\gamma^{\beta}\frac{1\pm\gamma_{5}}{2}u^{l}(k)\bar{u}^{\mu}(p')\gamma^{\nu}(q_{\mu}g_{\beta\nu}-q_{\nu}g_{\beta\mu})\frac{1\pm\gamma_{5}\slashed{s_{p}}}{2}u^{N}(p)&~~J^P=\frac{3}{2}^{-} RR,LL,\\
\frac{eg_{\gamma PP_c}}{m_{J/\psi}q^2}\bar{u}^{l'}(k')\gamma^{\beta}\frac{1\pm\gamma_{5}}{2}u^{l}(k)\bar{u}^{\mu}(p')\gamma^{\nu}(q_{\mu}g_{\beta\nu}-q_{\nu}g_{\beta\mu})\frac{1\mp\gamma_{5}\slashed{s_{p}}}{2}u^{N}(p)&~~J^P=\frac{3}{2}^{-} RL,LR.
\end{cases}
\label{eq:amplitudes_polized}
\end{align}

When calculating the polarized invariant amplitude, we use the
same coupling constants which were derived previously as only a
given initial  polarization state is taken. Note that among
the four possible combinations of electron and proton
polarizations, only two cases are independent as $RR$ and $ LL$, as
well as $RL$ and $LR$, result in the same invariant amplitudes.
In the result, the different handedness under the same parity changes the sign of $m_{p}$ term and $m_{P_{c}}$ terms, whereas the different parity under the same handedness changes the sign of $m_{P_{c}}$ term only.  More details are shown in Appendix \ref{appendix:polarized amplitude}.

\subsection{Cross section as functions of pseudorapidity and
transverse momentum}

We evaluate the differential cross section as functions of pseudorapidity ($\eta$) and transverse momentum (\pt).
$\eta$ is chosen as the main observable instead of rapidity because $\eta$ is directly connected to detector geometries in experiment.
Using the squared matrix amplitudes for $e+p$ scattering which is detaield in Appendix
\ref{appendix:unpolarized amplitude} and Appendix
\ref{appendix:polarized amplitude}, the differential cross section
in the CM frame is:

\begin{align}
&(\frac{d\sigma}{d\theta})_{CM}=\frac{2\pi\sin\theta}{64\pi^{2}
E_{CM}^{2}}\frac{|\vec{p_{f}}|}{|\vec{p_{i}}|}|\mathcal{M}|^{2}
\label{eqn:cross_section}
\end{align}

, where $|\vec{p_{i}}|=\frac{s-m_{p}^{2}}{2\sqrt{s}}$~and $|\vec{p_{f}}|=\frac{s-m_{P_{c}}^{2}}{2\sqrt{s}}$ are the initial and final
momentum in the CM frame, respectively.  
$\theta$ is the polar angle
of electrons after scattering in the CM frame.
After all, the 4-momentum of \pc~is boosted back to the lab frame to obtain the $\eta$-differential cross section.

\section{result}
In this section, we present the differential cross sections of \pc(4312) production in the $e+p$ collision at \sqrts = 126 GeV which is the EIC energy.  In accordance with the previous section, the results are studied as functions of $\eta$ and \pt~under four cases of $J^{P}=\frac{1}{2}^\pm$ and $\frac{3}{2}^\pm$.  
$\eta$ of \pc(4312) is computed in the lab frame, thus we can judge whether it arrives in the typical detector coverage proposed for the EIC ($|\eta|<4$). 
Fig.~\ref{fig:unpol} shows the differential cross sections for unpolarized $e+p$ collision. 

The numbers of \pc(4312)~ expected to be produced at the EIC with an integrated luminosity of $10~fb^{-1}$ is tabulated in Tab.~\ref{table:electro-production}.  This luminosity value, $10~fb^{-1}$, can be reached by running the EIC for about a month at the peak intensity ($10^{34}~cm^{-2}s^{-1}$), 8 hours a day.  We found that the  expected yields for the positive parity is larger than those for the negative parity by a factor 5, independent of \pt~and $\eta$.  The largest yield is expected for $J^{P}=\frac{3}{2}^{+}$.  Supposing a detector system  measures the \jpsi~via $e^{+}+e^{-}$ decay (branching ratio = 5.94\%) with the 100\% efficiency for electron and proton, $\mathcal{O}(10^{6})$ \pc 's are expected to be observed in the data accumulated for one month.


\begin{figure}[H]
\centering
\includegraphics[width=2.34in,height=1.71in]{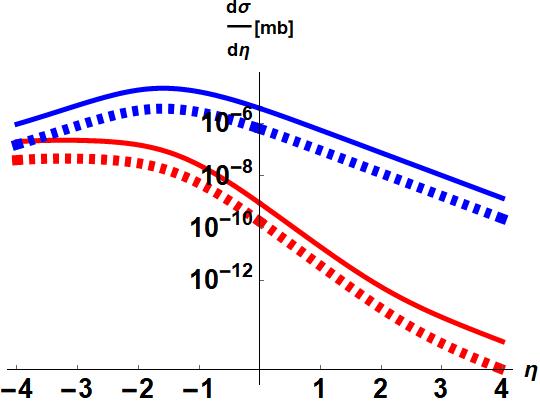}
\includegraphics[width=3.645in,height=1.71in]{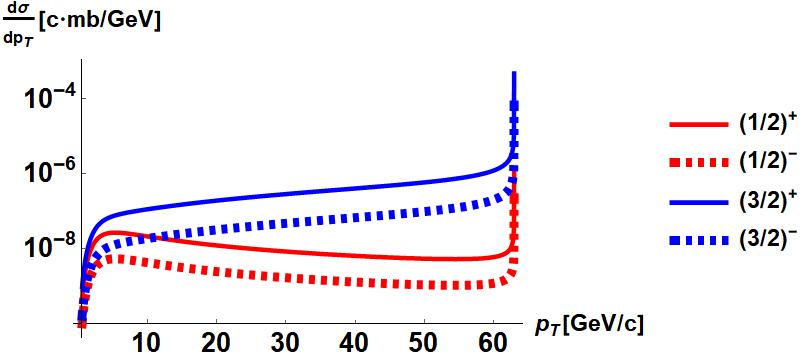}
\caption{Differential cross section of \pc ~production in the unpolarized $e+p$ collision for each case of spin-$\frac{1}{2}$ and spin-$\frac{3}{2}$ with positive and negative parity states. The results are calculated as a function of (a) $\eta$ and (b) \pt ($|\eta|<4$)}
\label{fig:unpol}
\end{figure}

\begin{table}[H]
\caption{Expected number of $P_{c}(4312)$ produced at the EIC with $10~fb^{-1}$.}
\centering
 \begin{tabular}{|| c | c | c | c | c ||}
 \hline
  $J^{P}$ of \pc~ & $\frac{1}{2}^{+}$   &   $\frac{1}{2}^{-}$   &   $\frac{3}{2}^{+}$   &   $\frac{3}{2}^{-}$    \\
   \hline
   Yield &  $~5.09\times10^{6}~$  &  $~1.01\times10^{6}~$  &  $~4.51\times10^{8}~$  &  $~7.46\times10^{7}~$   \\
  \hline
\end{tabular}
\label{table:electro-production}
\end{table}

\subsection{Polarized cross section}
The differential cross sections for the polarized electron and proton beams are shown in Fig.~\ref{fig:ptc}~(spin-$\frac{1}{2}$), and Fig.~\ref{fig:ptc2}~(spin-$\frac{3}{2}$).  In the case of spin-$\frac{1}{2}$, the cross sections of RR (same handedness) and RL (opposite handedness) configuration are almost identical for the backward rapidity region (proton-going direction), and they split in the forward region, $\eta >2$ (electron-going direction).  In the case of spin-$\frac{3}{2}$, a more dramatic behavior is observed: the cross section curves for RR and RL begin to separate early from $\eta \approx -2$, making RL cross section larger than RR one by two orders of magnitude at $\eta = 4$.  For clear observation of this effect in experiment, we propose to measure the forward-to-backward ratio (RFB) and the beam spin asymmetry (BSA), which are defined as follows. 

\begin{align}
&RFB~(\eta) = \frac{d\sigma/d\eta~(+\eta)}{d\sigma/d\eta~(-\eta)}, \mathrm{~~where}~\eta > 0 \\
&BSA~(\eta) = \frac{d\sigma/d\eta~[RL] - d\sigma/d\eta~(RR)}{d\sigma/d\eta~[RL] + d\sigma/d\eta~[RR]} 
\label{eq:rfb_bsa}
\end{align}

These observables have experimental benefit because some of uncertainties, such as luminosity, tracking correction, and geometric acceptance, are cancelled out.  As shown in Fig.~\ref{fig:bsa}, the spin of \pc~can be clearly determined by measuring the BSA in the mid-rapidity region.  Yet, we found that both BSA and RFB are not much useful to judge the parity. In particular, if \pc~ was in the spin-$\frac{3}{2}$ state, the BSA and RFB are completely insensitive to the parity.

\subsection{Determination of \pc's parity using $J/\psi$ polarization }

As shown above, it is hard to identify the parity of~\pc~ with only the cross section result. To cope with this problem, we further investigate the polarization of $J/\psi$. $J/\psi$ is a spin-1 massive vector boson with two transverse and one longitudinal polarization, thus having an anisotropic angular distribution for $J/\psi \rightarrow e^{+}+e^{-}$. 
The decay angle ($\theta$) is defined, in the rest frame of $J/\psi$, as the angle between the electron momentum and boost direction of the $J/\psi$ in the lab frame.
By measuring $\theta$, one can experimentally tune the transverse-to-longitudinal ratio as shown in Fig.~\ref{fig:polarization} (a).  After tagging the polarity of $J/\psi$, we study the dependence of matrix amplitude on $\phi$~which is defined as the decay angle of $J/\psi$ from \pc~in the rest frame of \pc. 

As shown in Fig.~\ref{fig:polarization}, the $\phi$~ distribution is significantly sensitive to the polarity of  $J/\psi$ for both spin-$\frac{1}{2}$ and spin-$\frac{3}{2}$ states.  In either cases, the difference between the transverse $J/\psi$ events (T) and the longitudinal ones (L) is more dramatic in the positive parity state than in the negative parity state. 
\begin{figure}[H]
\centering
\includegraphics[width=2.7in,height=1.85in]{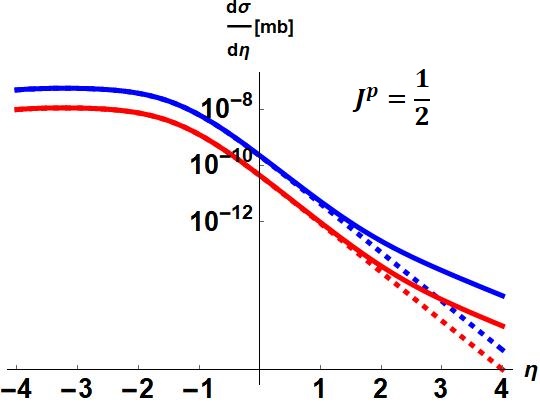}
\includegraphics[width=4.05in,height=1.75in]{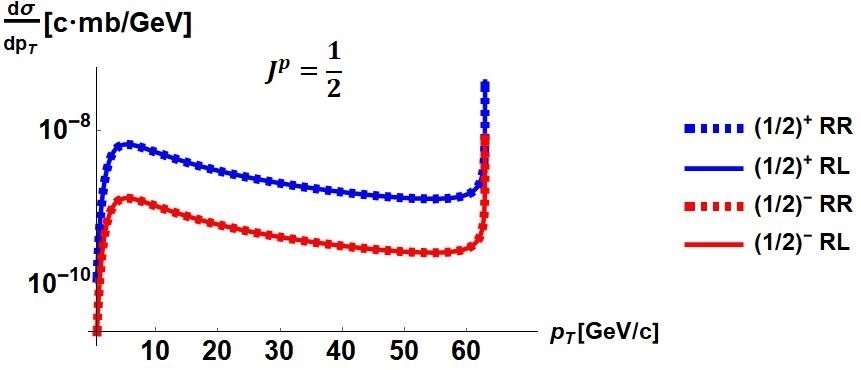}
\caption{The differential corss sections for spin$-\frac{1}{2}$ cases with polarized collision. R and L mean right-handed and left-handed, respectively  }
\label{fig:ptc}
\end{figure}

\begin{figure}[H]
\centering
\includegraphics[width=2.7in,height=1.85in]{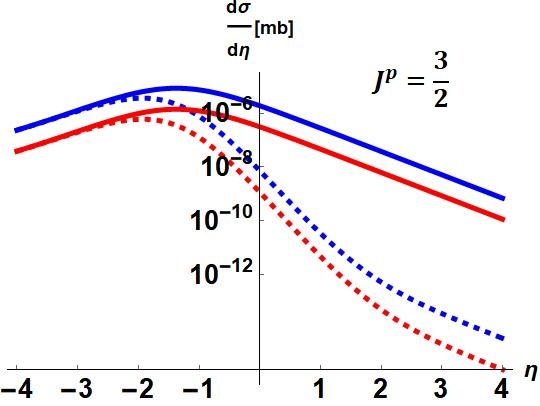}
\includegraphics[width=4.08in,height=1.85in]{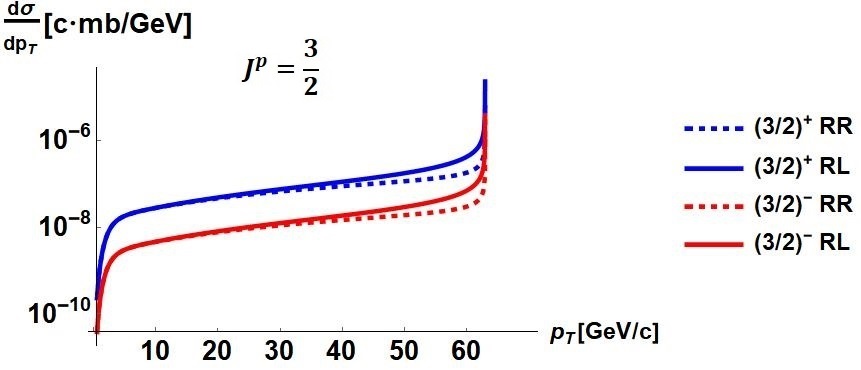}
\caption{The differential cross sections for spin-$\frac{3}{2}$ cases. R and L mean right-handed and left-handed, respectively}\label{fig:ptc2}
\end{figure}

\begin{figure}[H]
\centering
\includegraphics[width=3.4in,height=1.9in]{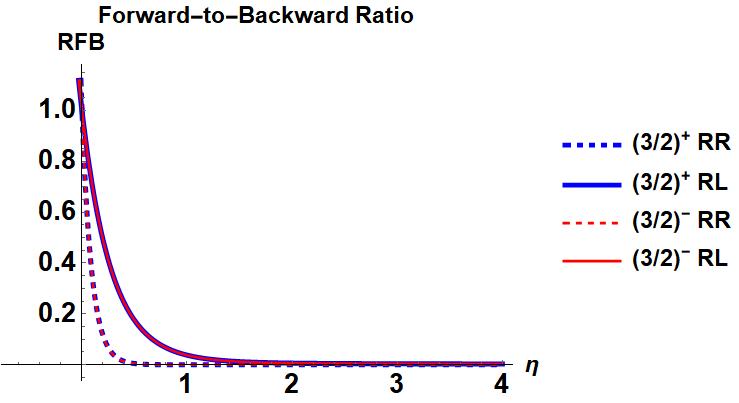}
\includegraphics[width=3.5in,height=1.9in]{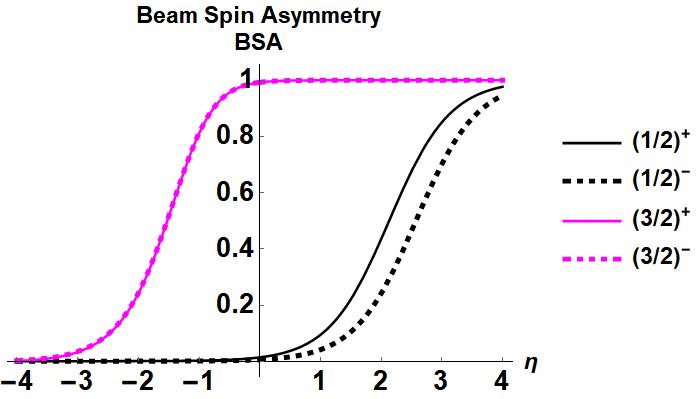}
\label{fig:ratio}
\caption{The forward-to-backward ratio (RFB) for spin-$\frac{3}{2}$ \pc~ state.
(b) Beam spin asymmetry (BSA) results for $J^{P}=\frac{1}{2}^\pm$ and $\frac{3}{2}^\pm$ states.}
\label{fig:bsa}
\end{figure}


\begin{figure}[H]
\includegraphics[width=2.2in,height=1.3in]{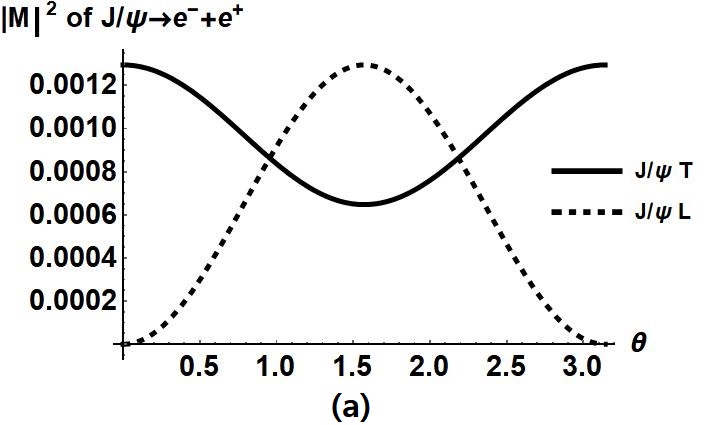}
\includegraphics[width=2.45in,height=1.3in]{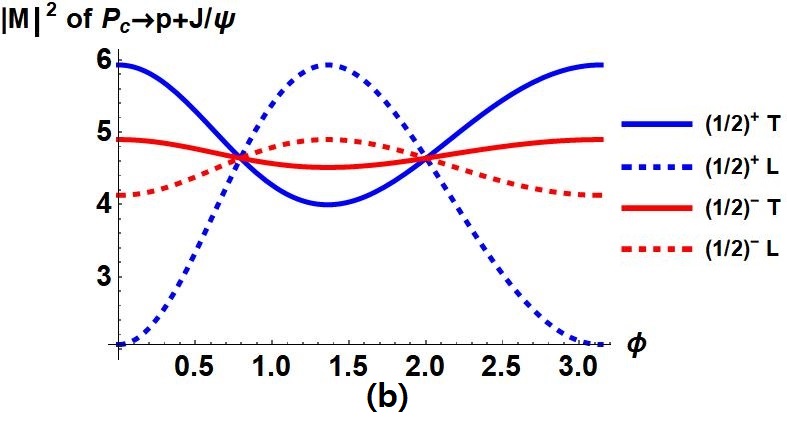}
\includegraphics[width=2.4in,height=1.3in]{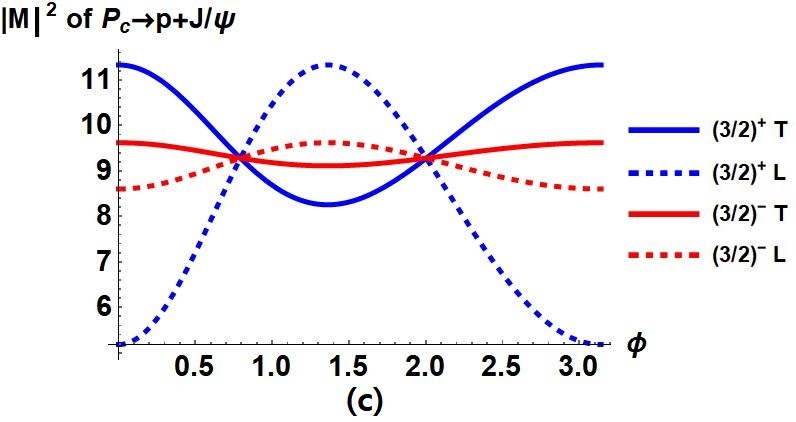}
\caption{(a)~$J/\psi\to e^{-}e^{+}$ amplitude as a function of $\theta$ for a  transverse [T] $J/\psi$ and a  longitudinal $J/\psi$ [L].  \\
~~~(b)~$(J_{P_c}=\frac{1}{2})\to p+J/\psi$ amplitude dependence on the decayed  $J/\psi$ polarization (T or L). \\
~~~(c)~$(J_{P_c}=\frac{3}{2})\to p+J/\psi$ amplitude dependence on the decayed  $J/\psi$ polarization (T or L).}
\label{fig:polarization}
\end{figure}
\end{widetext}

\section{Summary}
The cross section for the $P_c(4312)$ production in $e+p$ collision is studied under various assumptions for its potential quantum states; $J^P=\frac{1}{2}^{\pm}$ and $J^P=\frac{3}{2}^{\pm}$.

The interaction strength of the electro-production of $P_c(4312)$, created by scattering $\gamma$ onto a proton, is calculated using the vector meson dominance hypothesis to the leading order.  We also assume that the $P_c(4312)$ $\rightarrow$ \jpsi + $p$ channel is dominant in the decay width of $P_c(4312)$ that was measured by the LHCb collaboration. 
The cross section is larger for the spin-$\frac{3}{2}$ state than for the spin-$\frac{1}{2}$ state, and larger for the positive parity case than for the negative parity.  With one month of operation at the EIC in its nominal condition, millions of \pc(4312)'s are expected to be measured via $p+e^{+}+e^{-}$ channel.  This calculation can be generalized for other heavy pentaquarks as far as it can be electro-produced onto a proton.  
Furthermore, more kinds of pentaquarks can be produced by electro-production onto a neutron using $e+d$ collision at the EIC. Hence, the EIC can be considered as a factory of heavy pentaquarks and will provide an excellent opportunity for a comprehensive understanding of exotic particles. 

Given the availability of polarized beams at the EIC, we suggest that the analysis of pseudorapidity distribution of \pc~can confirm its spin number.  The forward-to-backward ratio and the beam-spin asymmetry results are unambiguously distinct for the spin-$\frac{1}{2}$ and spin-$\frac{3}{2}$ states.   These observables are also useful to reduce the experimental uncertainties as well.  

In addition, we prove that the decay kinematics of \pc $ \rightarrow p+$\jpsi ~is sensitive to the parity of \pc.  The distribution of the decay angle of \pc~depends on the polarization of the \jpsi, which can be statistically determined by measuring its decay angle of $e^{-}+e^{+}$. Therefore, the parity of \pc~ can be determined by the analysis of angular distribution. For this purpose, a hermetic detector with efficient calorimeters and tracking systems, such as ATHENA and ECCE, is necessary.

\section*{Acknowledgments}

This work was supported by Samsung Science and Technology Foundation under Project 
Number SSTF-BA1901-04, the POSCO Science Fellowship of POSCO TJ Park Foundation, 
and the National Research Foundation of Korea (NRF) of the Korea government
(MSIT) (No. 2018R1A5A1025563 and No. 2019R1A2C1087107).

\clearpage
\appendix
\begin{widetext}
\renewcommand{\thesection}{\Alph{section}}
\renewcommand{\thesubsection}{\Alph{section}-\arabic{subsection}}

\section{Invariant matrix amplitude for three-particles involving a $P_{c}$ }

\subsection{$P_{c}(p')\to J/\psi(q)+N(p)$}\label{appendix:jppc}

Here, we present invariant matrix elements for the decay of
the $P_c$ in all possible spin-parity states, i.e., four possible
states: $J^P=\frac{1}{2}^\pm, \frac{3}{2}^\pm$ using the
interaction Lagrangians given in Eq.~\eqref{eq:Lagrangian},

\begin{align}
|\mathcal{M}|^{2}=
\begin{cases}
\frac{g_{JpP_c}^2}{m_{J/\psi}^{2}}(-8(p\cdot p')m_{J/\psi}^{2}+32(q\cdot p)(q\cdot p')-24m_{p}m_{P_c}m_{J/\psi}^{2})\;& J^p=\frac{1}{2}^{+},\\\\
\frac{g_{JpP_c}^2}{m_{J/\psi}^{2}}(-8(p\cdot p')m_{J/\psi}^{2}+32(q\cdot p)(q\cdot p')+24m_{p}m_{P_c}m_{J/\psi}^{2})\;& J^p=\frac{1}{2}^{-},\\\\
\frac{2g_{JpP_c}^2}{3m_{J/\psi}^{2}}(\frac{2(q\cdot p')^2(p\cdot p')}{m_{P_c}^2}+2(q\cdot p)(q\cdot p')-m_{J/\psi}^{2}(p\cdot p'+3m_{p}m_{P_c}))\;& J^p=\frac{3}{2}^{+},\\\\
\frac{2g_{JpP_c}^2}{3m_{J/\psi}^{2}}(\frac{2(q\cdot p')^2(p\cdot
p')}{m_{P_c}^2}+2(q\cdot p)(q\cdot p')-m_{J/\psi}^{2}(p\cdot
p'-3m_{p}m_{P_c}))\;& J^p=\frac{3}{2}^{-}.
\end{cases}
\end{align}
where, $p\cdot
p'=\frac{1}{2}(m_{p}^{2}+m_{P_c}^{2}-m_{J/\psi}^{2}),\;q\cdot
p=\frac{1}{2}(m_{P_{c}}^{2}-m_{p}^{2}-m_{J/\psi}^{2}),\;q\cdot
p'=\frac{1}{2}(m_{P_c}^{2}+m_{J/\psi}^{2}-m_{p}^{2}),
p^{2}=m_{p}^{2},\;q^{2}=m_{J/\psi}^{2},\;p'^{2}=m_{P_c}^{2}$. The
decay rate is then given by,
\begin{align}
&\Gamma=\frac{1}{8\pi}\frac{|\vec{p_{f}}|}{E_{CM}^{2}}|\mathcal{M}|^{2},
\end{align}
where $\vec{p_{f}}$ is the momentum of $J/\psi$ and proton in the CM frame after decay of $P_c$,
\begin{align}\label{eqn:pf}
|\vec{p_{f}}|=\frac{1}{2m_{P_{c}}}\sqrt{(m_{P_{c}}^{2}-(m_{p}+m_{J/\psi})^{2})(m_{P_{c}}^{2}-(m_{p}-m_{J/\psi})^{2})}.
\end{align}

To find the 4-momentum and polarization vector of the $J/\psi$, we
take an inverse Lorentz transformation of them from the rest frame
of the $J/\psi$ to the Lab frame. In the $P_{c}$ decay, the
$P_{c}$ is boosted along the $z$-axis and the $J/\psi$ is boosted
along an arbitrary direction.

\begin{align}
\begin{cases}
q^{\mu}&=(m_{J/\psi},0,0,0)\to(q_{0},q_{1},q_{2},q_{3}),\nonumber\\
\varepsilon^{\mu}_{(1)}&=(0,1,0,0)\to\bigg(\frac{q_{1}}{m_{J/\psi}},1+\frac{q_{1}^{2}}{m_{J/\psi}(q_{0}+m_{J/\psi})},\frac{q_{1}q_{2}}{m_{J/\psi}(q_{0}+m_{J/\psi})},\frac{q_{1}q_{3}}{m_{J/\psi}(q_{0}+m_{J/\psi})}\bigg),\nonumber\\
\varepsilon^{\mu}_{(2)}&=(0,0,1,0)\to\bigg(\frac{q_{2}}{m_{J/\psi}},\frac{q_{1}q_{2}}{m_{J/\psi}(q_{0}+m_{J/\psi})},1+\frac{q_{2}^{2}}{m_{J/\psi}(q_{0}+m_{J/\psi})},\frac{q_{2}q_{3}}{m_{J/\psi}(q_{0}+m_{J/\psi})}\bigg),\nonumber\\
\varepsilon^{\mu}_{(3)}&=(0,0,0,1)\to\bigg(\frac{q_{3}}{m_{J/\psi}},\frac{q_{1}q_{3}}{m_{J/\psi}(q_{0}+m_{J/\psi})},\frac{q_{2}q_{3}}{m_{J/\psi}(q_{0}+m_{J/\psi})},1+\frac{q_{3}^{2}}{m_{J/\psi}(q_{0}+m_{J/\psi})}\bigg),
\end{cases}
\end{align}
where $q_{0}$ is the energy of the $J/\psi$, and $q_{i}(i=1,2,3)$
is the 3-momentum of the $J/\psi$. We adopt the Metric tensor,
$g_{00}=1,\;g_{0i}=g_{i0}=0,\;g_{ij}=-\delta_{ij}$.

In order to distinguish the difference between positive and
negative spin-parity states, we exhibit transverse and
longitudinal parts of the matrix amplitude by using
\begin{align}
&P_{\mu\nu}=\sum_{a=1}^{3}\varepsilon^{(a)}_{\mu}\varepsilon^{(a)}_{\nu}
=\varepsilon^{T}_{\mu}\varepsilon^{T}_{\nu}+\varepsilon^{L}_{\mu}
\varepsilon^{L}_{\nu}=P^{T}_{\mu\nu}+P^{L}_{\mu\nu}=-g_{\mu\nu}+
\frac{q_{\mu}q_{\nu}}{m_{J/\psi}^{2}},
\end{align}
where superscripts, $T$ and $L$ stand for transverse and
longitudinal directions, respectively. Using the 4-momentum and
polarization vectors given above, we can get transverse and
longitudinal part of the polarization tensor,
\begin{align}
&P_{\mu\nu}^{T}=\begin{pmatrix}0&0\\0&\delta_{ij}-\frac{q_{i}q_{j}}
{\vec{q}^{2}}\end{pmatrix},\;P_{\mu\nu}^{L}=\begin{pmatrix}
\frac{\vec{q}^{2}}{m_{J/\psi}^{2}}&-\frac{q_{0}q_{i}}{m_{J/\psi}^{2}}\\
-\frac{q_{0}q_{i}}{m_{J/\psi}^{2}}&\frac{q_{0}^{2}q_{i}q_{j}}{m_{J/\psi}^{2}\vec{q}^{2}}
\end{pmatrix}
\end{align}
and matrix amplitude.

\begin{align}
&J^{p}=\frac{1}{2}^{+}
\begin{cases}
|\mathcal{M}|^{2}_{T}&=\frac{32g_{JpP_c}^{2}}{m_{J/\psi}^{2}}\bigg(2(q\cdot p)(q\cdot p')+\frac{m_{J/\psi}^{2}}{\vec{q}^{2}}(\vec{q}\cdot\vec{p})(\vec{q}\cdot\vec{p'})-m_{J/\psi}^{2}(E_{\vec{p}}E_{\vec{p'}}+m_{p}m_{P_c})\bigg), \\
|\mathcal{M}|^{2}_{L}&=\frac{16g_{JpP_c}^{2}}{m_{J/\psi}^{2}}\bigg(-m_{J/\psi}^{2}(p\cdot p'+m_{p}m_{P_c})+2m_{J/\psi}^{2}E_{\vec{p}}E_{\vec{p'}}-\frac{2m_{J/\psi}^{2}}{\vec{q}^{2}}(\vec{q}\cdot\vec{p})(\vec{q}\cdot\vec{p'})\bigg),\\
\end{cases}\\
&J^{p}=\frac{1}{2}^{-}
\begin{cases}
|\mathcal{M}|^{2}_{T}&=\frac{32g_{JpP_c}^{2}}{m_{J/\psi}^{2}}\bigg(2(q\cdot p)(q\cdot p')+\frac{m_{J/\psi}^{2}}{\vec{q}^{2}}(\vec{q}\cdot\vec{p})(\vec{q}\cdot\vec{p'})-m_{J/\psi}^{2}(E_{\vec{p}}E_{\vec{p'}}-m_{p}m_{P_c})\bigg), \\
|\mathcal{M}|^{2}_{L}&=\frac{16g_{JpP_c}^{2}}{m_{J/\psi}^{2}}\bigg(-m_{J/\psi}^{2}(p\cdot p'-m_{p}m_{P_c})+2m_{J/\psi}^{2}E_{\vec{p}}E_{\vec{p'}}-\frac{2m_{J/\psi}^{2}}{\vec{q}^{2}}(\vec{q}\cdot\vec{p})(\vec{q}\cdot\vec{p'})\bigg), \\
\end{cases}\\
&J^{p}=\frac{3}{2}^{+}
\begin{cases}
|\mathcal{M}|^{2}_{T}&=\frac{8g_{JpP_{c}}^{2}}{3m_{P_c}^{2}m_{J/\psi}^{2}}\bigg(2m_{P_c}^{2}(q\cdot p)(q\cdot p')+2(p\cdot p')(q\cdot p')^{2}-2m_{p}m_{P_c}^{3}m_{J/\psi}^{2}-m_{P_c}^{2}(p\cdot p')m_{J/\psi}^{2}  \\
&+\frac{m_{J/\psi}^{2}(p\cdot p')}{\vec{q}^{2}}(\vec{q}\cdot\vec{p'})^{2}+\frac{m_{J/\psi}^{2}}{\vec{q}^{2}}m_{P_c}^{2}(\vec{q}\cdot\vec{p})(\vec{q}\cdot\vec{p'})-m_{J/\psi}^{2}\vec{p'}^{2}(p\cdot p')-m_{P_c}^{2}E_{\vec{p}}E_{\vec{p'}}m_{J/\psi}^{2}\bigg), \\
|\mathcal{M}|^{2}_{L}&=\frac{8g_{JpP_{c}}^{2}}{3m_{P_c}^{2}m_{J/\psi}^{2}}\bigg(-m_{p}m_{P_c}^{3}m_{J/\psi}^{2}-m_{P_c}^{2}(\vec{q}\cdot\vec{p})(\vec{q}\cdot\vec{p'})\frac{m_{J/\psi}^{2}}{\vec{q}^{2}}-(p\cdot p')(\vec{q}\cdot\vec{p'})^{2}\frac{m_{J/\psi}^{2}}{\vec{q}^{2}} \\
&+m_{J/\psi}^{2}\vec{p'}^{2}(p\cdot
p')+m_{P_c}^{2}E_{\vec{p}}E_{\vec{p'}}m_{J/\psi}^{2}\bigg),
\end{cases}\\
&J^{p}=\frac{3}{2}^{-}
\begin{cases}
|\mathcal{M}|^{2}_{T}&=\frac{8g_{JpP_{c}}^{2}}{3m_{P_c}^{2}m_{J/\psi}^{2}}\bigg(2m_{P_c}^{2}(q\cdot p)(q\cdot p')+2(p\cdot p')(q\cdot p')^{2}+2m_{p}m_{P_c}^{3}m_{J/\psi}^{2}-m_{P_c}^{2}(p\cdot p')m_{J/\psi}^{2} \\
&+\frac{m_{J/\psi}^{2}(p\cdot p')}{\vec{q}^{2}}(\vec{q}\cdot\vec{p'})^{2}+\frac{m_{J/\psi}^{2}}{\vec{q}^{2}}m_{P_c}^{2}(\vec{q}\cdot\vec{p})(\vec{q}\cdot\vec{p'})-m_{J/\psi}^{2}\vec{p'}^{2}(p\cdot p')-m_{P_c}^{2}E_{\vec{p}}E_{\vec{p'}}m_{J/\psi}^{2}\bigg), \\
|\mathcal{M}|^{2}_{L}&=\frac{8g_{JpP_{c}}^{2}}{3m_{P_c}^{2}m_{J/\psi}^{2}}\bigg(m_{p}m_{P_c}^{3}m_{J/\psi}^{2}-m_{P_c}^{2}(\vec{q}\cdot\vec{p})(\vec{q}\cdot\vec{p'})\frac{m_{J/\psi}^{2}}{\vec{q}^{2}}-(p\cdot p')(\vec{q}\cdot\vec{p'})^{2}\frac{m_{J/\psi}^{2}}{\vec{q}^{2}} \\
&+m_{J/\psi}^{2}\vec{p'}^{2}(p\cdot
p')+m_{P_c}^{2}E_{\vec{p}}E_{\vec{p'}}m_{J/\psi}^{2}\bigg).
\end{cases}
\end{align}

We obtain the 4-momentum of particles by performing an inverse
Lorentz transformation from the CM frame (rest frame of the
$P_{c}$) to the Lab frame. The result is given by,
\begin{align}\label{eqn:4momentum_Pc}
&p'^{\mu}=(E_{\vec{p'}},\vec{p'})=(\gamma m_{P_c},0,0,\gamma m_{P_c}\beta), \nonumber \\
&p^{\mu}=(E_{\vec{p}},\vec{p})=\bigg(\gamma(\sqrt{\vec{p_{f}}^{2}+m_{p}^{2}}+\beta|\vec{p_{f}}|\cos\phi),|\vec{p_{f}}|\sin\phi,0,\gamma(|\vec{p_{f}}|\cos\phi+\beta\sqrt{\vec{p_{f}}^{2}+m_{p}^{2}})\bigg) \nonumber \\
&q^{\mu}=(E_{\vec{q}},\vec{q})=\bigg(\gamma(\sqrt{\vec{p_{f}}^{2}+m_{J/\psi}^{2}}-\beta|\vec{p_{f}}|\cos\phi),-|\vec{p_{f}}|\sin\phi,0,\gamma(-|\vec{p_{f}}|\cos\phi+\beta\sqrt{\vec{p_{f}}^{2}+m_{J/\psi}^{2}})\bigg).
\end{align}
with $|\vec{p}_{f}|$ being the momentum defined in
({\ref{eqn:pf}}), and $\phi$ being the polar angle of the proton
in the CM frame with respect to boost axis of the $P_{c}$.

\subsection{$J/\psi\to\gamma\to e^{-}+e^{+}$}\label{appendix:jpsi}

The invariant matrix amplitude of the $J/\psi$ decaying into a
positron and an electron is given as below.
\begin{align}
&\mathcal{M}=(-i)\frac{e^{2}q^{2}}{g_{J}}\varepsilon^{\mu}_{J}\frac{-i(g_{\mu\nu}-\frac{q_{\mu}q_{\nu}}{m_{J/\psi}^{2}})}{q^{2}}\bar{u}_{e^{-}}\gamma^{\nu}v_{e^{+}}=-\frac{e^{2}}{g_{J}}\varepsilon_{J}^{\mu}\bar{u}_{e^{-}}\gamma_{\mu}v_{e^{+}}.
\end{align}

Averaging over initial polarization of the J/$\psi$, we
get,

\begin{align}
|\mathcal{M}|^{2}&=\frac{64\pi^{2}\alpha^{2}}{3g_{J}^{2}}(m_{J/\psi}^{2}+2m_{l}^{2}).
\end{align}
with $m_{l}$ being the electron mass and $\alpha=e^2/4\pi$ is a fine structure constant. Using the method in Appendix \ref{appendix:jppc}, we can also obtain the transverse and longitudinal matrix amplitude of $J/\psi(q)\to e^{-}(k)+e^{+}(k')$.

\begin{align}
|\mathcal{M}|^{2}_{T}
&=\frac{128\pi^{2}\alpha^{2}}{g_{J}^{2}}\bigg(E_{\vec{k}}E_{\vec{k'}}+m_{l}^{2}-\frac{(\vec{q}\cdot\vec{k})(\vec{q}\cdot\vec{k'})}{\vec{q}^{2}}\bigg),\nonumber\\
|\mathcal{M}|^{2}_{L}&=\frac{64\pi^{2}\alpha^{2}}{g_{J}^{2}}\bigg(k\cdot
k'+m_{l}^{2}+\frac{2E_{\vec{k}}E_{\vec{k'}}\vec{q}^{2}}{m_{J/\psi}^{2}}+\frac{2E_{\vec{q}}^{2}(\vec{q}\cdot\vec{k})(\vec{q}\cdot\vec{k'})}{\vec{q}^{2}m_{J/\psi}^{2}}-\frac{2E_{\vec{q}}E_{\vec{k}}(\vec{q}\cdot\vec{k'})}{m_{J/\psi}^{2}}-\frac{2E_{\vec{q}}E_{\vec{k'}}(\vec{q}\cdot\vec{k})}{m_{J/\psi}^{2}}\bigg).
\end{align}
Similarly as shown in Eq(\ref{eqn:4momentum_Pc}), we obtain the
momentum in the Lab frame by taking an inverse Lorentz
transformation from the CM frame (rest frame of the $J/\psi$) to
the Lab frame,
\begin{align}
&q^{\mu}=(E_{\vec{q}},\vec{q})=(\gamma m_{J/\psi},0,0,\gamma
m_{J/\psi}\beta), \nonumber \\
&k^{\mu}=(E_{\vec{k}},\vec{k})=\bigg(\gamma(E_{\vec{p}_{f}}+\beta|\vec{p}_{f}|\cos\theta),|\vec{p}_{f}|\sin\theta,0,\gamma(|\vec{p}_{f}|\cos\theta+\beta E_{\vec{p}_{f}})\bigg),\nonumber\\
&k'^{\mu}=(E_{\vec{k'}},\vec{k'})=\bigg(\gamma(E_{\vec{p}_{f}}-\beta|\vec{p}_{f}|\cos\theta),-|\vec{p}_{f}|\sin\theta,0,\gamma(-|\vec{p}_{f}|\cos\theta+\beta
E_{\vec{p}_{f}})\bigg)
\end{align}
with the magnitude of momentum in the CM frame,
$|\vec{p_{f}}|=\frac{1}{2}\sqrt{m_{J/\psi}^{2}-4m_{l}^{2}}$ and
$E_{\vec{p}_{f}}=\sqrt{\vec{p}_{f}^{2}+m_{l}^{2}}$. $\theta$ is
the polar angle of an electron in the CM frame with respect to the
boost axis of the $J/\psi$. Then, the decay rate becomes,
\begin{equation}
\Gamma=\frac{4\pi}{3}(\frac{\alpha}{g_{J}})^{2}\sqrt{m_{J/\psi}^{2}
-4m_{l}^{2}}(1+\frac{2m_{l}^{2}}{m_{J/\psi}^{2}}),
\end{equation}
and from the decay rate, we obtain $g_{J}=11.2$.

\section{Invariant matrix amplitudes for the scattering between unpolarized electrons and protons}\label{appendix:unpolarized amplitude}

Here, we present the square of the invariant matrix elements shown
in Eq. (\ref{eq:amplitudes}) for the electro-production of the
$P_c$ in all possible four spin-parity states,
$J^P=\frac{1}{2}^\pm, \frac{3}{2}^\pm$ obtained from the
interaction Lagrangians given in Eq.~\eqref{eq:Lagrangian}. The
invariant matrix amplitude square for the scattering between
unpolarized electrons and protons are as follows.

\begin{eqnarray}
|\mathcal{M}|^2_{J^p=\frac{1}{2}^{+}}&=&\frac{64\pi\alpha g_{\gamma pP_c}^2}{m_{J/\psi}^2q^4}\bigg(-2(k\cdot p)(k'\cdot
p')q^2-2(k'\cdot p)(k\cdot p')q^2+(k\cdot k')(p\cdot
p')q^2+2(k\cdot p)(q\cdot k')(q\cdot p')
\nonumber \\
&+&2(k\cdot p')(q\cdot k')(q\cdot p)+2(k'\cdot p')(q\cdot
k)(q\cdot p)+2(k'\cdot p)(q\cdot k)(q\cdot p')-2(p\cdot p')(q\cdot
k)(q\cdot k') \nonumber \\
&-&2m_{p}m_{P_c}(q\cdot k)(q\cdot k')-m_{p}m_{P_c}(k\cdot
k')q^2\bigg),\\
|\mathcal{M}|^2_{J^p=\frac{1}{2}^{-}}&=&\frac{64\pi\alpha g_{\gamma pP_c}^2}{m_{J/\psi}^2q^4}\bigg( -2(k\cdot p)(k'\cdot
p')q^2-2(k'\cdot p)(k\cdot p') q^2+(k\cdot k') (p\cdot
p')q^2+2(k\cdot p)(q\cdot k')(q\cdot p')
\nonumber \\
&+&2(k\cdot p')(q\cdot k')(q\cdot p)+2(k'\cdot p')(q\cdot k)
(q\cdot p)+2(k'\cdot p)(q\cdot k)(q\cdot p')-2(p\cdot p')
(q\cdot k)(q\cdot k') \nonumber \\
&+&2m_{p}m_{P_c}(q\cdot k)(q\cdot k')+m_{p}m_{P_c}(k\cdot
k')q^2\bigg),
\label{half_polized}
\end{eqnarray}
Similarly
for the $P_c$ with its spin $\frac{3}{2}$,

\begin{eqnarray}
|\mathcal{M}|^2_{J^p=\frac{3}{2}^{+}}&=&\frac{32\pi\alpha g_{\gamma pP_{c}}^{2}}{3m_{P_{c}}^{2}m_{J/\psi}^2q^{4}}\bigg(-2m_{p}m_{P_c}^3(q\cdot
k)(q\cdot k') -m_{P_c}^2q^{2}(k\cdot k')(m_{p}m_{P_c}-p\cdot p')
-2m_{P_c}^2(q\cdot k)(p\cdot p')(q\cdot k') \nonumber \\
&+&m_{P_c}^2(q\cdot k)(k'\cdot p)(q\cdot p') +m_{P_c}^2(k\cdot
p)(q\cdot k')(q\cdot p')+m_{P_c}^2(q\cdot k) (q\cdot p)(k'\cdot
p')-m_{P_c}^2q^2(k\cdot p)(k'\cdot p')
\nonumber \\
&-&(k\cdot p')\bigg(m_{P_c}^2q^2(k'\cdot p)-m_{P_c}^2 (q\cdot
p)(q\cdot k')+2(p\cdot p')(q^2(k'\cdot p')-(q\cdot k')
(q\cdot p'))\bigg) \nonumber \\
&+&2(q\cdot k)(p\cdot p')(k'\cdot p')(q\cdot p')\bigg), \\
|\mathcal{M}|^2_{J^p=\frac{3}{2}^{-}}&=&\frac{32\pi\alpha g_{\gamma pP_{c}}^{2}}{3m_{P_{c}}^{2}m_{J/\psi}^2q^{4}}\bigg(2m_{p}m_{P_c}^3(q\cdot
k)(q\cdot k') +m_{P_c}^2q^{2}(k\cdot k')(m_{p}m_{P_c}+p\cdot p')
-2m_{P_c}^2(q\cdot k)(p\cdot p')(q\cdot k') \nonumber \\
&+&m_{P_c}^2(q\cdot k)(k'\cdot p)(q\cdot p')+m_{P_c}^2 (k\cdot
p)(q\cdot k')(q\cdot p')+m_{P_c}^2(q\cdot k)(q\cdot p)
(k'\cdot p')-m_{P_c}^2q^2(k\cdot p)(k'\cdot p') \nonumber \\
&+&(k\cdot p')\bigg(-m_{P_c}^2q^2(k'\cdot p) +m_{P_c}^2(q\cdot
p)(q\cdot k')+2(p\cdot p')((q\cdot k')(q\cdot
p')-q^2(k'\cdot p'))\bigg) \nonumber \\
&+&2(q\cdot k)(p\cdot p')(k'\cdot p')(q\cdot p')\bigg).
\label{threehalves_polized}
\end{eqnarray}
with $k$ being the
momentum of incoming electrons, $p$ being the momentum of incoming
protons, $k'$ being the momentum of outgoing electrons, and $p'$
being the momentum of outgoing $P_{c}$, $q=k-k'=p'-p$. In the CM frame, those momenta can be written in
terms of $|\vec{p}_{i}|$, $|\vec{p}_{f}|$ and $\theta$ defined in
Eq.(\ref{eqn:cross_section}).
\begin{align}
&k^{\mu}=(|\vec{p}_{i}|,0,0,|\vec{p}_{i}|), \nonumber \\
&p^{\mu}=(\sqrt{\vec{p}_{i}^{2}+m_{p}^{2}},0,0,-|\vec{p}_{i}|), \nonumber \\
&k'^{\mu}=(|\vec{p}_{f}|,|\vec{p}_{f}|\sin\theta,0,|\vec{p}_{f}|\cos\theta),
\nonumber \\
&p'^{\mu}=(\sqrt{\vec{p}_{f}^{2}+m_{P_c}^{2}},-|\vec{p}_{f}|\sin\theta,0,-|\vec{p}_{f}|\cos\theta).
\end{align}

\section{Invariant matrix amplitudes for the scattering between
polarized electrons and protons}\label{appendix:polarized amplitude}

Here, we present the invariant matrix
elements for the scattering between polarized electrons and
protons shown in Eq. (\ref{eq:amplitudes_polized}). We use the
notation $s_{p}$ to refer to the 4-spin vector of the proton. The absolute value square of
invariant matrix elements for polarized electrons and protons are
given by,

{\allowdisplaybreaks
\begin{eqnarray}
|\mathcal{M}|^2_{J^p=\frac{1}{2}^{+} RR,LL}&=&\frac{16\pi\alpha g_{\gamma pP_c}^2}{m_{J/\psi}^2q^4}\bigg(q^2(k\cdot k')(p\cdot p'-m_{p}
m_{P_c})-2\Big((q\cdot k)\big((q\cdot k')(p\cdot p'+m_{p}m_{P_c})
+m_{p}(s_{p}\cdot k')(q\cdot p') \nonumber \\
&+& m_{p}(q\cdot s_{p})(k'\cdot p')+(k'\cdot p)(m_{P_c}(q\cdot
s_{p}) -q\cdot p')-m_{P_c}(q\cdot p)(s_{p}\cdot k')-(q\cdot
p)(k'\cdot p')\big) \nonumber \\
&-&m_{p}(s_{p}\cdot k)(q\cdot k')(q\cdot p')-(k\cdot p')
\big((q\cdot k')(m_{p}(q\cdot s_{p})+q\cdot p)-q^{2}(k'\cdot
p)\big)+m_{P_c}(s_{p}\cdot k)(q\cdot p)(q\cdot k') \nonumber \\
&-&m_{P_c}(k\cdot p)(q\cdot s_{p})(q\cdot k')-(k\cdot p)(q\cdot
k') (q\cdot p')+q^{2}(k\cdot p)(k'\cdot p')\Big)\bigg), \\
|\mathcal{M}|^2_{J^p=\frac{1}{2}^{+} RL,LR}&=&\frac{16\pi\alpha g_{\gamma PP_c}^2}{m_{J/\psi}^2q^4}\bigg(q^2(k\cdot k')(p\cdot p'
-m_{p}m_{P_c})-2\Big((q\cdot k)\big((q\cdot k')(p\cdot p'
+m_{p}m_{P_c})-m_{p}(s_{p}\cdot k')(q\cdot p') \nonumber \\
&-&m_{p}(q\cdot s_{p})(k'\cdot p')-(k'\cdot p)(m_{P_c}(q\cdot
s_{p})+q\cdot p')+m_{P_c}(q\cdot p)(s_{p}\cdot k')-(q\cdot
p)(k'\cdot p')\big) \nonumber \\
&+& m_{p}(s_{p}\cdot k)(q\cdot k')(q\cdot p')+(k\cdot p')
\big((q\cdot k')(m_{p}(q\cdot s_{p})-q\cdot p)+q^{2}(k'\cdot
p)\big)-m_{P_c}(s_{p}\cdot k)(q\cdot p)(q\cdot k') \nonumber \\
&+& m_{P_c}(k\cdot p)(q\cdot s_{p})(q\cdot k')-(k\cdot p)(q\cdot
k')(q\cdot p')+q^{2}(k\cdot p)(k'\cdot p')\Big)\bigg),
\end{eqnarray} }
and for the tensor coupling in the negative parity
$J^p=\frac{1}{2}^{-}$,

{\allowdisplaybreaks
\begin{eqnarray}
|\mathcal{M}|^2_{J^p=\frac{1}{2}^{-} RR,LL}&=&\frac{16\pi\alpha g_{\gamma pP_c}^2}{m_{J/\psi}^2q^4}\bigg(q^2(k\cdot k')(p\cdot p'
+m_{p}m_{P_c})+2\Big((q\cdot k)\big((q\cdot k')(m_{p}m_{P_c}
-p\cdot p')-m_{p}(s_{p}\cdot k')(q\cdot p') \nonumber \\
&-& m_{p}(q\cdot s_{p})(k'\cdot p')+(k'\cdot p)(m_{P_c} (q\cdot
s_{p})+q\cdot p')-m_{P_c}(q\cdot p)(s_{p}\cdot k')+(q\cdot
p)(k'\cdot p')\big) \nonumber \\
&+& m_{p}(s_{p}\cdot k)(q\cdot k')(q\cdot p')+(k\cdot p')
\big((q\cdot k')(m_{p}(q\cdot s_{p})+q\cdot p)-q^{2}(k'\cdot
p)\big)+m_{P_c}(s_{p}\cdot k)(q\cdot p)(q\cdot k') \nonumber \\
&-& m_{P_c}(k\cdot p)(q\cdot s_{p})(q\cdot k')+(k\cdot p)(q\cdot
k')(q\cdot p')-q^{2}(k\cdot p)(k'\cdot p')\Big)\bigg), \\
|\mathcal{M}|^2_{J^p=\frac{1}{2}^{-} RL,LR}&=&\frac{16\pi\alpha g_{\gamma pP_c}^2}{m_{J/\psi}^2q^4}\bigg(q^2(k\cdot k')(p\cdot p'
+m_{p}m_{P_c})+2\Big((q\cdot k)\big((q\cdot k')(m_{p}m_{P_c}
-p\cdot p')+m_{p}(s_{p}\cdot k')(q\cdot p') \nonumber \\
&+& m_{p}(q\cdot s_{p})(k'\cdot p')+(k'\cdot p)(q\cdot p'
-m_{P_c}(q\cdot s_{p}))+m_{P_c}(q\cdot p)(s_{p}\cdot k') +(q\cdot
p)(k'\cdot p')\big) \nonumber \\
&-& m_{p}(s_{p}\cdot k)(q\cdot k')(q\cdot p')-(k\cdot p')
\big((q\cdot k')(m_{p}(q\cdot s_{p})-q\cdot p)+q^{2}(k'\cdot p)\big)
-m_{P_c}(s_{p}\cdot k)(q\cdot p)(q\cdot k') \nonumber \\
&+& m_{P_c}(k\cdot p)(q\cdot s_{p})(q\cdot k')+(k\cdot p)(q\cdot
k')(q\cdot p')-q^{2}(k\cdot p)(k'\cdot p')\Big)\bigg).
\end{eqnarray} }
Similarly for the coupling in the positive parity
$J^p=\frac{3}{2}^{+}$,

{\allowdisplaybreaks
\begin{eqnarray}
|\mathcal{M}|^2_{J^p=\frac{3}{2}^{+} RR,LL}&=&-\frac{8\pi\alpha g_{\gamma pP_{c}}^{2}}{3m_{P_{c}}^{2}m_{J/\psi}^2q^{4}} \bigg(2m_{p}m_{P_c}^{3}(q\cdot k)(q\cdot
k')+m_{p}m_{P_c}^{2}(s_{p} \cdot k)(q\cdot k')(q\cdot p')
-m_{p}m_{P_c}^{2}(q\cdot k)(s_{p}\cdot k')(q\cdot p') \nonumber \\
&-& m_{p}m_{P_c}^{2}(q\cdot k)(q\cdot s_{p})(k'\cdot p')+(k\cdot
p')\Big(m_{p}m_{P_c}^{2}(q\cdot s_{p})(q\cdot k')+2m_{p}(q\cdot
k')(q\cdot p')(s_{p}\cdot p') \nonumber \\
&+& m_{P_c}^{2}q^{2}(k'\cdot p)+2m_{P_c}(p\cdot p')(q\cdot
s_{p})(q\cdot k')-m_{P_c}(q\cdot p) (q\cdot k')(2(s_{p}\cdot
p')+m_{P_c})-2(p\cdot p')(q\cdot k')(q\cdot p')
\nonumber \\
&+& 2q^{2}(p\cdot p')(k'\cdot p')\Big)+m_{P_c}^{2}q^{2}(k\cdot k')
(m_{p} m_{P_c}-p\cdot p')-2m_{p}(q\cdot k)(k'\cdot p')(q\cdot p')(s_{p}
\cdot p') \nonumber \\
&+& m_{P_c}^{3}(q\cdot k)(q\cdot s_{p})(k'\cdot p)+m_{P_c}^{3}
(s_{p}\cdot k)(q\cdot p)(q\cdot k')-m_{P_c}^{3} (k\cdot p)(q\cdot
s_{p})(q\cdot k')-m_{P_c}^{3}(q\cdot k)(q\cdot p)(s_{p}\cdot k')
\nonumber \\
&+& 2m_{P_c}^{2}(q\cdot k)(p\cdot p')(q\cdot k')-m_{P_c}^{2}
(q\cdot k)(k'\cdot p)(q\cdot p')-m_{P_c}^{2}(k\cdot p)(q\cdot
k')(q\cdot p')-m_{P_c}^{2}(q\cdot k) (q\cdot p)(k'\cdot p')
\nonumber \\
&+& m_{P_c}^{2}q^{2}(k\cdot p)(k'\cdot p')-2m_{P_c}(s_{p}\cdot k)
(p\cdot p')(q\cdot k')(q\cdot p')+2m_{P_c}(q\cdot k)(p\cdot
p')(s_{p}\cdot k')(q\cdot p') \nonumber \\
&-& 2m_{P_c} (q\cdot k)(k'\cdot p)(q\cdot p')(s_{p}\cdot p')
+2m_{P_c}(k\cdot p)(q\cdot k')(q\cdot p')(s_{p}\cdot p')-2m_{P_c}
(q\cdot k)(p\cdot p')(q\cdot s_{p})(k'\cdot p') \nonumber \\
&+& 2m_{P_c}(q\cdot k)(q\cdot p)(k'\cdot p')(s_{p}\cdot p')
-2(q\cdot k)(p\cdot p')(k'\cdot p')(q\cdot p')\bigg), \\
|\mathcal{M}|^2_{J^p=\frac{3}{2}^{+} RL,LR}&=&-\frac{8\pi\alpha g_{\gamma pP_{c}}^{2}}{3m_{P_{c}}^{2}m_{J/\psi}^2q^{4}} \bigg(2m_{p}m_{P_c}^{3}(q\cdot k)(q\cdot
k')-m_{p}m_{P_c}^{2}(s_{p} \cdot k)(q\cdot k')(q\cdot p')
+m_{p}m_{P_c}^{2}(q\cdot k)(s_{p}\cdot k')(q\cdot p') \nonumber \\
&+& m_{p}m_{P_c}^{2}(q\cdot k)(q\cdot s_{p})(k'\cdot p')+(k\cdot
p')\Big(-m_{p}m_{P_c}^{2}(q\cdot s_{p})(q\cdot k')-2m_{p}(q\cdot
k')(q\cdot p')(s_{p}\cdot p') \nonumber \\
&+& m_{P_c}^{2}q^{2}(k'\cdot p)-2m_{P_c}(p\cdot p')(q\cdot
s_{p})(q\cdot k')-m_{P_c}(q\cdot p) (q\cdot k')(-2(s_{p}\cdot
p')+m_{P_c})-2(p\cdot p')(q\cdot k')(q\cdot p')
\nonumber \\
&+& 2q^{2}(p\cdot p')(k'\cdot p')\Big)+m_{P_c}^{2}q^{2}(k\cdot k')
(m_{p} m_{P_c}-p\cdot p')+2m_{p}(q\cdot k)(k'\cdot p')(q\cdot
p')(s_{p}
\cdot p') \nonumber \\
&-& m_{P_c}^{3}(q\cdot k)(q\cdot s_{p})(k'\cdot p)-m_{P_c}^{3}
(s_{p}\cdot k)(q\cdot p)(q\cdot k')+m_{P_c}^{3} (k\cdot p)(q\cdot
s_{p})(q\cdot k')+m_{P_c}^{3}(q\cdot k)(q\cdot p)(s_{p}\cdot k')
\nonumber \\
&+& 2m_{P_c}^{2}(q\cdot k)(p\cdot p')(q\cdot k')-m_{P_c}^{2}
(q\cdot k)(k'\cdot p)(q\cdot p')-m_{P_c}^{2}(k\cdot p)(q\cdot
k')(q\cdot p')-m_{P_c}^{2}(q\cdot k) (q\cdot p)(k'\cdot p')
\nonumber \\
&+& m_{P_c}^{2}q^{2}(k\cdot p)(k'\cdot p')+2m_{P_c}(s_{p}\cdot k)
(p\cdot p')(q\cdot k')(q\cdot p')-2m_{P_c}(q\cdot k)(p\cdot
p')(s_{p}\cdot k')(q\cdot p') \nonumber \\
&+& 2m_{P_c} (q\cdot k)(k'\cdot p)(q\cdot p')(s_{p}\cdot p')
-2m_{P_c}(k\cdot p)(q\cdot k')(q\cdot p')(s_{p}\cdot p')+2m_{P_c}
(q\cdot k)(p\cdot p')(q\cdot s_{p})(k'\cdot p') \nonumber \\
&-& 2m_{P_c}(q\cdot k)(q\cdot p)(k'\cdot p')(s_{p}\cdot p')
-2(q\cdot k)(p\cdot p')(k'\cdot p')(q\cdot p')\bigg),
\end{eqnarray} }
and, for the coupling in the negative parity
$J^p=\frac{3}{2}^{-}$,

{\allowdisplaybreaks
\begin{eqnarray}
|\mathcal{M}|^2_{J^p=\frac{3}{2}^{-} RR,LL}&=&\frac{8\pi\alpha g_{\gamma pP_{c}}^{2}}{3m_{P_{c}}^{2}m_{J/\psi}^2q^{4}} \bigg(2m_{p}m_{P_c}^{3}(q\cdot k)(q\cdot
k')-m_{p}m_{P_c}^{2}(s_{p} \cdot k)(q\cdot k')(q\cdot p')
+m_{p}m_{P_c}^{2}(q\cdot k)(s_{p}\cdot k')(q\cdot p') \nonumber \\
&+& m_{p}m_{P_c}^{2}(q\cdot k)(q\cdot s_{p})(k'\cdot p')-(k\cdot
p')\Big(m_{p}m_{P_c}^{2}(q\cdot s_{p})(q\cdot k')+2m_{p}(q\cdot
k')(q\cdot p')(s_{p}\cdot p') \nonumber \\
&+& m_{P_c}^{2}q^{2}(k'\cdot p)-2m_{P_c}(p\cdot p')(q\cdot
s_{p})(q\cdot k')-m_{P_c}(q\cdot p) (q\cdot k')(-2(s_{p}\cdot
p')+m_{P_c})-2(p\cdot p')(q\cdot k')(q\cdot p')
\nonumber \\
&+& 2q^{2}(p\cdot p')(k'\cdot p')\Big)+m_{P_c}^{2}q^{2}(k\cdot k')
(m_{p} m_{P_c}+p\cdot p')+2m_{p}(q\cdot k)(k'\cdot p')(q\cdot
p')(s_{p}
\cdot p') \nonumber \\
&+& m_{P_c}^{3}(q\cdot k)(q\cdot s_{p})(k'\cdot p)+m_{P_c}^{3}
(s_{p}\cdot k)(q\cdot p)(q\cdot k')-m_{P_c}^{3} (k\cdot p)(q\cdot
s_{p})(q\cdot k')-m_{P_c}^{3}(q\cdot k)(q\cdot p)(s_{p}\cdot k')
\nonumber \\
&-& 2m_{P_c}^{2}(q\cdot k)(p\cdot p')(q\cdot k')+m_{P_c}^{2}
(q\cdot k)(k'\cdot p)(q\cdot p')+m_{P_c}^{2}(k\cdot p)(q\cdot
k')(q\cdot p')+m_{P_c}^{2}(q\cdot k) (q\cdot p)(k'\cdot p')
\nonumber \\
&-& m_{P_c}^{2}q^{2}(k\cdot p)(k'\cdot p')-2m_{P_c}(s_{p}\cdot k)
(p\cdot p')(q\cdot k')(q\cdot p')+2m_{P_c}(q\cdot k)(p\cdot
p')(s_{p}\cdot k')(q\cdot p') \nonumber \\
&-& 2m_{P_c} (q\cdot k)(k'\cdot p)(q\cdot p')(s_{p}\cdot p')
+2m_{P_c}(k\cdot p)(q\cdot k')(q\cdot p')(s_{p}\cdot p')-2m_{P_c}
(q\cdot k)(p\cdot p')(q\cdot s_{p})(k'\cdot p') \nonumber \\
&+& 2m_{P_c}(q\cdot k)(q\cdot p)(k'\cdot p')(s_{p}\cdot p')
+2(q\cdot k)(p\cdot p')(k'\cdot p')(q\cdot p')\bigg), \\
|\mathcal{M}|^2_{J^p=\frac{3}{2}^{-} RL,LR}&=&\frac{8\pi\alpha g_{\gamma pP_{c}}^{2}}{3m_{P_{c}}^{2}m_{J/\psi}^2q^{4}} \bigg(2m_{p}m_{P_c}^{3}(q\cdot k)(q\cdot
k')+m_{p}m_{P_c}^{2}(s_{p} \cdot k)(q\cdot k')(q\cdot p')
-m_{p}m_{P_c}^{2}(q\cdot k)(s_{p}\cdot k')(q\cdot p') \nonumber \\
&-& m_{p}m_{P_c}^{2}(q\cdot k)(q\cdot s_{p})(k'\cdot p')+(k\cdot
p')\Big(m_{p}m_{P_c}^{2}(q\cdot s_{p})(q\cdot k')+2m_{p}(q\cdot
k')(q\cdot p')(s_{p}\cdot p') \nonumber \\
&-& m_{P_c}^{2}q^{2}(k'\cdot p)-2m_{P_c}(p\cdot p')(q\cdot
s_{p})(q\cdot k')+m_{P_c}(q\cdot p) (q\cdot k')(2(s_{p}\cdot
p')+m_{P_c})+2(p\cdot p')(q\cdot k')(q\cdot p')
\nonumber \\
&-& 2q^{2}(p\cdot p')(k'\cdot p')\Big)+m_{P_c}^{2}q^{2}(k\cdot k')
(m_{p} m_{P_c}+p\cdot p')-2m_{p}(q\cdot k)(k'\cdot p')(q\cdot
p')(s_{p}
\cdot p') \nonumber \\
&-& m_{P_c}^{3}(q\cdot k)(q\cdot s_{p})(k'\cdot p)-m_{P_c}^{3}
(s_{p}\cdot k)(q\cdot p)(q\cdot k')+m_{P_c}^{3} (k\cdot p)(q\cdot
s_{p})(q\cdot k')+m_{P_c}^{3}(q\cdot k)(q\cdot p)(s_{p}\cdot k')
\nonumber \\
&-& 2m_{P_c}^{2}(q\cdot k)(p\cdot p')(q\cdot k')+m_{P_c}^{2}
(q\cdot k)(k'\cdot p)(q\cdot p')+m_{P_c}^{2}(k\cdot p)(q\cdot
k')(q\cdot p')+m_{P_c}^{2}(q\cdot k) (q\cdot p)(k'\cdot p')
\nonumber \\
&-& m_{P_c}^{2}q^{2}(k\cdot p)(k'\cdot p')+2m_{P_c}(s_{p}\cdot k)
(p\cdot p')(q\cdot k')(q\cdot p')-2m_{P_c}(q\cdot k)(p\cdot
p')(s_{p}\cdot k')(q\cdot p') \nonumber \\
&+& 2m_{P_c} (q\cdot k)(k'\cdot p)(q\cdot p')(s_{p}\cdot p')
-2m_{P_c}(k\cdot p)(q\cdot k')(q\cdot p')(s_{p}\cdot p')+2m_{P_c}
(q\cdot k)(p\cdot p')(q\cdot s_{p})(k'\cdot p') \nonumber \\
&-& 2m_{P_c}(q\cdot k)(q\cdot p)(k'\cdot p')(s_{p}\cdot p')
+2(q\cdot k)(p\cdot p')(k'\cdot p')(q\cdot p')\bigg).
\end{eqnarray} }

\end{widetext}

\pagebreak

\end{document}